\documentstyle[prl,aps,epsf]{revtex}\def\narrowtext{}\tighten\twocolumn
\begin{document}
\draft

\title{
Cu Nuclear Quadrupole Resonance Study of 
La$_{2-x}$Sr$_x$Cu$_{1-y}$Zn$_y$O$_4$ ($x$=0.10, 0.15 and 0.20):
Zn-induced Wipeout Effect 
near the Magnetic and Electric Instability
}
\author{
H. Yamagata$^1$, H. Miyamoto$^1$, K. Nakamura$^1$, 
M. Matsumura$^1$, and Y. Itoh$^{2,*}$
}

\address{
$^1$Department of Material Science, Faculty of Science, Kochi University, Kochi 780-8520, Japan\\ 
$^2$Japan Society for the Promotion of Science, Tokyo, Japan\\
}

%\date{\today}%
\date{November 21, 2001}%
\maketitle %

%\address{\begin{minipage}[t]{6.0in} %
\begin{abstract}%
We studied Zn-substitution effect on the high-$T_c$ superconductors, underdoped La$_{2-x}$Sr$_x$Cu$_{1-y}$Zn$_y$O$_4$ ($x$=0.10;
$y$=0, 0.01, 0.02), optimally doped ($x$=0.15; $y$=0, 0.02), and overdoped ($x$=0.20; $y$=0, 0.03, 0.06) in a temperature range of
$T$=4.2-300 K, using Cu nuclear quadrupole resonance (NQR) spin-echo technique. 
We found full disappearance of the Cu NQR signals for
the Zn-substituted, Sr-underdoped $x$=0.10 samples below about 40 K, partial disappearance 
for the Sr-optimally doped ones below about 50 K, 
but not for the overdoped $x$=0.20 ones. 
From the Zn-doping, the Sr-doping and the temperature dependence of the wipeout effect, 
we associate the wipeout effect with Zn-induced Curie magnetism or its extended glassy charge-spin stripe formation. 
%\typeout{polish abstract} %
\end{abstract}
%\end{minipage}} %

%\maketitle %
\narrowtext

\section{introduction}

	 Nonmagnetic Zn$^{2+}$ ion has a closed shell of the 3$d$ electrons, so that it must not carry a local moment. However, 
the actual Zn
substitution strongly suppresses the transition temperature $T_c$ of underdoped high-$T_c$ superconductors~\cite{Xiao0}, 
induces a Curie-like uniform susceptibility~\cite{Xiao1,Xiao2,Harashina,Monod,Alloul,Mendels}, 
and finally causes a quasi-elastic magnetic correlation~\cite{Hirota}. 
In contrast to a simple dilution effect, Yamagata, Inada and Matsumura discovered a significant loss
of zero-field Cu nuclear quadrupole resonance (NQR) spectrum intensity (wipeout effect) for the Zn-impurity doped
YBa$_2$Cu$_3$O$_{7-\delta}$~\cite{Yamagata}. 
Subsequent muon spin relaxation study observed a Zn-induced magnetic fluctuation with a
short relaxation time ($<$ 10 $\mu$s) for the Zn-doped YBa$_2$Cu$_3$O$_y$ ($y$=6.43-6.67)~\cite{Mendels1}, 
and inelastic neutron scattering study
observed the Zn-induced low frequency spin fluctuations around an antiferromagnetic wave vector 
for the Zn-doped YBa$_2$Cu$_3$O$_{6.97}$~\cite{Harashina1,Sidis}.  
Much attention has been paid for the novel microscopic effect of Zn doping~\cite{Walstedt,Mahajan,Itoh}.

Here, the wipeout effect is defined as disappearance of the observable NMR or NQR signal 
around a rf frequency region in a longer time scale than a few $\mu$s. 
For the nuclear spin system in a material, the temperature
dependence of the integrated NMR/NQR signal should follow a Curie law, 
because the nuclear spins in aÊthermal equilibrium state follow
the Boltzman distribution. 
However, imperfection or impurity changes a spatial environment 
around the nuclei place by place, so that the
NMR/NQR spectrum is broadened or split into several pieces. 
The nuclei around a magnetic impurity feel a strong local field from the
impurity, so that several neighbor nuclei surrounding the impurity are unobservable. 
The strong wipeout effects have been observed for
the lightly doped insulators~\cite{Matsumura}
and the Co-doped high-$T_c$ superconductors~\cite{Matsumura1}. 
From careful measurements of the Zn-doping dependence
of the integrated Cu(2) NQR spectrum for YBa$_2$Cu$_3$O$_{7-\delta}$, 
the first neighbor shell surrounding the Zn-occupied Cu(2) site 
was inferred to be unobservable~\cite{Yamagata}. 
Similar observation is reported in refs.~\cite{Walstedt1,Walstedt}. 
The strong wipeout effect has also been observed for the underdoped
La$_{2-x}$Sr$_x$CuO$_4$ ($x\leq$0.125), which is associated with a charge-stripe ordering or its glassy nature
~\cite{Hunt,Hunt1,Julien}.   

In this paper, we report Cu NQR spectrum measurements for Zn-doped La$_{2-x}$Sr$_x$Cu$_{1-y}$Zn$_y$O$_4$ 
with (Sr-underdoped $x$=0.10; $y$=0, 0.01, 0.02), 
(optimally Sr-doped $x$=0.15; $y$=0, 0.02), and (Sr-overdoped $x$=0.20; $y$=0, 0.03, 0.06) 
in a temperature range of $T$=4.2-300 K. 
We found that an enhanced wipeout effect due to the Zn doping occurs in the Sr-underdoped samples 
but not for the overdoped ones. For the
Zn-doped $x$=0.10 sample, the Cu NQR signal completely diminishes below about 40 K. 
The diminished NQR signal recovers from the
Sr-underdoped to the overdoped regimes. 
From the Sr-doping dependence of the Zn-induced wipeout effect and the temperature dependence of
the wipeout effect from the normal to the superconducting states, 
we discuss the origin of the Zn-enhanced wipeout effect. 

\section{Experiments}

Polycrystalline samples were synthesized by a conventional solid-state-reaction method 
from CuO, ZnO, SrCO$_3$, and La$_2$O$_3$, as described in
refs.~\cite{Yoshimura,Itoh1}. 
The $X$-ray diffraction patterns indicated single phases. 
Nearly the same amount of samples (about 1.8 g of fine powders
with grain radii less than about 20 mm), so as to include nearly the same number of the Cu sites, 
were prepared for the NQR
measurements. All the samples for Cu NQR measurement were coated by paraffin oil. 
$T_c$ was determined from polycrystalline resistivity. 
In Fig. 1, $T_c$ is plotted as a function of Zn content $y$ for Sr doping of $x$=0.10. 0.15 and 0.20. 
The error bars are estimated from the onset
of resistivity drop and zero resistivity. 
$\Delta T_c/\Delta (100y)$ is estimated to be -16 K/\%
for $x$=0.10, -13 K/\% for $x$=0.15, and -7 K/\% for $x$=0.20,
which are similar to the reported values~\cite{Harashina1}. 
Zn doping suppresses $T_c$ of 
the underdoped materials stronger than of the overdoped ones.
The critical impurity concentration $y_c$ ($T_c$ $<$ 4.2 K) is estimated to be about 0.019 
for $x$=0.10 and about 0.042 for $x$=0.20.  

A home-made, phase-coherent-type pulsed spectrometer was utilized to do Cu NQR experiments. 
The Cu NQR frequency spectra with quadrature
detection were measured by integration of Cu nuclear spin-echoes 
with changing the frequency point by point in a two-pulse sequence
($\pi$/2-$\tau$-$\pi$-echo). 
Transverse relaxation curves of the Cu nuclear spin-echo $E(\tau)$ were also measured by the two-pulse method. 
A typical
width of the first exciting $\pi$/2-pulse $t_W$ was about 1 $\mu$s 
(the pulse strength $\nu_1\sim$250 kHz from 2$\pi\nu_1t_W$=$\pi$/2, 
or $H_1\sim$128 G from Ã3$\gamma_nH_1t_W$=$\pi$/2
with the $^{63}$Cu nuclear gyromagnetic ratio $\gamma_n$), 
the time interval $\tau$ for the NQR spectrum measurement was about 8$\sim$12 $\mu$s, and the number of
averaging times of spin-echos was about 1000. 

\section{Experimental Results}

Figure 2 shows the temperature dependence of the Cu NQR spectra of 
the Sr-underdoped La$_{1.9}$Sr$_{0.1}$Cu$_{1-y}$Zn$_y$O$_4$ with Zn doping of $y$=0, 0.01,
0.02 (left) and of the Sr-overdoped La$_{1.8}$Sr$_{0.2}$Cu$_{1-y}$Zn$_y$O$_4$ 
with Zn doping of $y$=0, 0.03, 0.06 (right), 
at $T$=77 K (top), 30 or 35 K (middle), and 4.2 K (bottom). 
The observed Cu NQR spectra for Zn-free materials agree with the reported ones~\cite{Yoshimura1,Yoshimura2}, 
which are understood by two sets (A and B) of $^{63, 65}$Cu NQR lines 
(the natural abundance ratio of $^{63}$Cu/$^{65}$Cu=0.69/0.31, and the resonance frequency
ratio of $^{63}\nu_Q/^{65}\nu_Q$= 0.211/0.195). 
The site assignment to A and B lines is given in refs.~\cite{Yoshimura1,Yoshimura2}. 
Obviously, Zn causes a strong wipeout
effect for the underdoped materials at lower temperatures. 
Although some decrease of the Cu NQR spectrum has been reported for the
Zn-free $x$=0.10~\cite{Hunt}, we found that the Cu NQR signals completely vanish at 4.2 K 
both for the Zn-doped but finite $T_c$ sample ($T_c$=13 K,
$y$=0.01) and for the Zn-doped, non-superconducting one ($T_c$ $<$ 4.2 K, $y$=0.02). 
However, Zn does not seem to affect the Cu NQR spectrum of
the overdoped ones. Thus, Zn effect is quite different between the underdoped and the overdoped regimes. 

Figure 3 shows the temperature dependence of transverse relaxation rates, Lorentzian decay rates 
1/$T_{2L}$ (closed circles) and Gaussian decay rates 1/$T_{2G}$ (open triangles), 
of the $^{63}$Cu(A) nuclear spin-echo decays (a) for the Zn-free $x$=0.10 
and (b) for the Zn-doped $x$=0.10 ($y$=0.02), and (c) for the Zn-free $x$=0.20  
and (d) for the Zn-doped $x$=0.20 ($y$=0.06). 
The relaxation times $T_{2L}$ and $T_{2G}$ were obtained from the least-squares fitting of a Gaussian-times-Lorentzian function 
$E(\tau)=E$(0)exp[-0.5(2$\tau/T_{2G})^2-(2\tau/T_{2L}$)]
($E$(0), $T_{2L}$ and $T_{2G}$ are the fitting parameters) to the experimental decay curves $E(\tau)$. 
The Lorentzian component is
predominant at low temperatures, which agrees with the previous similar report in the wipeout effect~\cite{Hunt}. 
The magnitudes of 1/$T_{2L}$ and
1/$T_{2G}$ increase with decreasing Sr content. 
A decrease of 1/$T_{2L}$ at $T_c$ in (a) and (c) starts at the onset of superconductivity
both for the Zn-free $x$=0.10 and 0.20 ($T_c \sim$29 K). 
As the temperature decreases, the significant loss of the Cu NQR spectra occurs
as in Fig. 2; nevertheless any strong divergence of the transverse relaxation rates was not observed for the Zn-doped, Sr-underdoped
sample in (b). 
For the Zn-doped, Sr-overdoped one in (d), an increase of 1/$T_{2L}$ was observed down to 4.2 K below the Zn-free
$T_c\sim$29 K. 

Figure 4 shows a typical separation of an observed Cu NQR spectrum of 
the ($x$, $y$)=(0.15, 0.02) sample at 4.2 K into two sets (A and B)
of $^{63}, ^{65}$Cu NQR lines by Gaussian distribution functions (solid curves).
 The fitting procedure is the same as those in refs.~\cite{Yoshimura1,Yoshimura2}. 

In Fig. 5, the estimated full-width at half maximum (FWHM) of $^{63}$Cu(A) NQR line is shown 
for $x$=0.10 ($y$=0, 0.01, 0.02) (left) and for $x$=0.20 ($y$=0. 0.03, 0.06) (right). 
The estimated FWHM is about  $\sim$2.0 MHz for $x$=0.10 and $\sim$2.2 MHz for $x$=0.20 
above $T$=100, which seem to be independent of the Zn content $y$. 
At lower temperatures than about 100 K, the linewidth is slightly broadened to be $\sim$2.3 MHz for both $x$.

Figure 6 shows the temperature dependence of the integrated intensity multiplied by temperature 
$I(x$=0.10, $y$), $I(x$=0.15, $y$) and $I(x$=0.2,$y$) for the Zn-free and Zn-doped samples. 
Figure 7 shows the ratio of the integrated Cu NQR spectra, $I(x$=0.10, $y$=0.01, 0.02)/$I(x$=0.10, $y$=0), 
$I(x$=0.15, $y$=0.02)/$I(x$=0.15, $y$=0), 
and $I(x$=0.20, $y$=0.03, 0.06)/$I(x$=0.20, $y$=0) (from the top to the bottom) as functions of temperature. 
Each integrated Cu NQR spectrum $I(x$, $y$) was corrected by a Boltzmann factor (1/$T$) 
and was extrapolated to $\tau\rightarrow0$ by using the
Gaussian-times-Lorentzian function with the decay times in Fig. 3 
[for the ($x$, $y$)=(0.15, 0.02) sample, the decay curve at 4.2 K was
reproduced by a two-exponential function better than the Gaussian-times-Lorentzian, 
so that only $I(x$=0.15, $y$=0.02) at 4.2 K was
extrapolated by a two-exponential function].

As shown in Fig. 6, for the Zn-free $x$=0.10 sample, 
the Cu NQR spectrum decreases monotonically below 50 K to 4.2 K,
 being consistent with the result in ref. [19]. 
This behavior is not observed for the Zn-free optimally doped $x$=0.15 and overdoped $x$=0.20
samples, whose Cu NQR signals increase with following a Boltzmann factor. 
In a higher temperature range of $T$=100-300 K, about 20 to 30 \%
of the Cu NQR signal is wiped out with substituting Zn impurities for $x$=0.10 and $x$=0.15 samples, 
while for overdoped ones ($x$=0.20) the
strong wipeout is not observed in a whole temperature range. 
By using the intensity ratio, one can observe a relative intensity loss of
the Cu NQR spectrum at each temperature as shown in Fig. 7. 
For the Zn-doped underdoped $x$=0.10 samples, the Cu NQR spectrum completely
disappears below about 40 K. 
For the Zn-doped optimal $x$=0.15 samples, the wipeout effect is incomplete even at 4.2 K. 
For the Zn-doped
overdoped $x$=0.20 samples, any loss of the signal intensity more than the site substitution is not observed.

\section{Discussions}
\subsection{Sr-underdoping-induced wipeout effect and Zn-induced wipeout effect}

The strong wipeout effect has already been observed for the Zn-free, deeply Sr-underdoped  La$_{2-x}$Sr$_x$CuO$_4$ (1/16 $< x <$ 1/8), 
which is associated with the charge-spin stripe ordering~\cite{Hunt} or 
the glassy slowing-down charge-spin fluctuations with distributed spin
fluctuation energy constants~\cite{Hunt1}. 
The charge-spin stripe ordering is theoretically proposed for doped Mott insulators~\cite{Zaane}, 
which is experimentally confirmed to be stabilized for a specific composition~\cite{Tranquada,Tranquada1}.

The Zn-induced wipeout effect, additionally for $x$=0.10 but newly for $x$=0.15, 
is not due to a simple lattice disorder, because any large
broadening of the line width or change of the line shape is not observed.
 For $x$=0.15, the temperature dependent wipeout effect is
obviously observed with Zn doping, which does not exist in the pure ($x$=0.15, $y$=0). 
Thus, the wipeout effect for optimally doped samples
of $x$=0.15 is a purely Zn-induced wipeout effect.

When the charge-spin stripe ordering is {\it static} at low temperatures, 
one must observe some pattern in a Zeeman-perturbed Cu NQR spectrum,
broadened by an internal magnetic field up to aboout 85 MHz, 
as has actually been observed for a specific composition, i.e. La$_{2-x}$Ba$_x$CuO$_4$
with $x\sim$1/8~\cite{Matsumura2,Hunt1}, 
La$_{2-x-y}$Sr$_x$Y$_y$CuO$_4$ with $x$=1/8~\cite{Matsumura3}, 
and La$_{2-x-y}$Eu$_y$Sr$_x$CuO$_4$~\cite{Sawa,Teitel'baum,Hunt1}. 
For the Zn-doped samples, however, we did not
observe any escaped signal at lower or higher frequency regions than those in Fig. 2 down to 1.3 K. 
We also did not observe any large
change of the line shape nor the line width as in Figs. 2 and 5. 
Thus, even if the stripe fluctuations exhibit slowing down behavior at
40 K, it must keep the {\it quasi-static} nature down to 1.3 K. 
The temperature dependence of Cu NQR intensity could not tell us whether the
wipeout sites make a stripe pattern or a random distribution on the CuO$_2$ plane in real space. 
Direct spatial information is lack in the
Cu NQR intensity.

From previous works~\cite{Hunt,Hunt1}, it is shown that there is strong wipeout effect 
in an extensive Sr concentration region (1/16 $< x <$ 1/8)
without Zn impurities. 
Then, the Zn substitution effect may help to extend the wipeout region up to $x$=0.15 and to increase the wipeout
fraction more than the samples of $y$=0. 
This is similar to Nd- or Eu-codoping effect. 
For Nd- or Eu-codoped samples, the optimal $T_c$ with
respect to $x$ is suppressed, the specific $T_c$ suppression region around $x$=1/8 is broadened, 
and then the wipeout region is extended into
the overdoping region up to $x$=0.20~\cite{Hunt1}. 
For Zn-substituted samples, the superconductivity itself is also suppressed, and the wipeout
region is also extended up to $x$=0.15. 
However, one should notice that there is a difference between the Sr-underdoping effect and the
Zn-induced effect. 
The Zn impurities induce Curie magnetism but not the Sr underdoping~\cite{Xiao1,Xiao2,Harashina,Monod}. 
The Zn doping readily induces a Curie-like
uniform susceptibility for underdoped YBa$_2$Cu$_{3-x}$Zn$_x$O$_{7-\delta}$, 
which is assigned to local moments induced by Zn through $^{89}$Y NMR study~\cite{Mahajan}.  
The Zn impurity induces satellites of $^{89}$Y NMR spectrum, which is not thought of the stripe pattern as in the $x$=1/8
systems~\cite{Matsumura2,Matsumura3,Sawa,Teitel'baum,Hunt1}. 
 The Curie-like uniform susceptibility is also observed 
for La$_{2-x}$Sr$_x$Cu$_{1-y}$Zn$_y$O$_4$~\cite{Xiao1,Xiao2,Harashina,Monod}. 
The observed strong wipeout effect due to Zn doping
closely resembles that due to dilute magnetic alloys~\cite{Nagasawa}. 
Thus, it is likely that Zn induces local moments near Zn and causes the strong wipeout effect in La$_{2-x}$Sr$_x$CuO$_4$. 
The Zn-induced Curie magnetism may extend a glassy state, similarly to the glassy stripe fluctuations. 
These results are sharply contrast to what is expected from a simple dilution effect, 
a dilute potential scattering for a Fermi liquid or normal residual density of states in a pair-breaking theory~\cite{Hotta}.

\subsection{Wipeout effect without precursor}

The abrupt wipeout effect without any precursor is one of the characteristics of the Zn-doping effect. 
The Cu NQR signal vanishes fully
or partially without divergence of the linewidth nor of 1/$T_2$ 
around the onset temperature $T_{NQR}$(=40$\sim$60 K) of the wipeout effect. 
Unobservable nuclei suddenly increase at $T_{NQR}$. 
This anomaly is not a conventional second order magnetic long-range ordering. 
One may speculate that the low temperature electronic state 
on the CuO$_2$ plane is microscopically and spatially segregated, 
so as to make it impossible to transfer information of electronic states from the unobserved sites to the observed ones. 
Such a segregated electronic
state being close to an antiferromagnetic instability is theoretically possible, 
if the superconductivity-to-antiferromagnetism
transition is of first order~\cite{Kohno}. 
The Mott transition is believed to be of first order. 

Let us illustrate schematically a model on spin fluctuations in Fig. 8, 
through the time spectrum of a muon spin relaxation $M(t)$ (a)
and the dynamical spin susceptibility $\chi "(Q_{in}, \omega$) as a function of frequency $\omega$ 
($Q_{in}$ is an incommensurate wave vector) (b), which are
inferred from the wipeout effect without any precursor below $T_{NQR}$. 
The arrows indicate the respective changes when cooling down below $T_{NQR}$. 
In Fig. 8(a), the dashed curves are extrapolated from longer relaxation components.
 Figure 8(a) shows the appearance of a fast
relaxation component ($<$ 5 $\mu$s) at lower temperatures. 
In Fig. 8(b), the shaded area is an observable NQR frequency region. 
The spin fluctuations peaked at a low frequency ($\sim$ 30 MHz) and 
at a high frequency ($\geq$10$^6$ MHz) in Fig. 8(b) correspond to fast and slow
relaxation components in Fig. 8(a), respectively. 
The present Cu NQR technique cannot detect such a fast relaxation component ($<$ 5 $\mu$s) in Fig. 8(a). 
  
This model is qualitatively evidenced by recent muon spin relaxation ($\mu$SR)~\cite{Panagopoulos} 
and inelastic neutron experiments~\cite{Kimura}.
 The systematic study by muon spin relaxation for La$_{2-x}$Sr$_x$Cu$_{1-y}$Zn$_y$O$_4$ 
with extensive ($x, y$)-compositions indicates the
existence of two characteristic temperatures $T_f$ and $T_g$ ($T_f > T_g$)~\cite{Panagopoulos}. 
With cooling down, a slowing down component in the spin
fluctuations grows, first enters the $\mu$SR time window at $T_f$ and then freezes into a glassy state at $T_g$. 
These characteristic temperatures depend on the hole and Zn concentrations 
and go to zero at $x$=0.2 irrespective of Zn concentration. 
Up to $x$=0.2 both $T_f$ and $T_g$ increase with increasing Zn concentration. 
After refs.~\cite{Hunt,Hunt1}, considering the difference in the time scales between NQR and $\mu$SR, 
our results are consistent with $\mu$SR; 
the observed full wipeout effect on Cu NQR for $x$=0.10 agrees with appearance of the short relaxation
component with a dominant fraction below $T_{NQR}\sim T_f$. 
Recent inelastic neutron scattering study for the optimally Sr-doped 
La$_{1.85}$Sr$_{0.15}$Cu$_{1-y}$Zn$_y$O$_4$, 
indicates that a small amount of Zn impurities strongly modifies the dynamical spin
susceptibility $\chi "(\omega$); 
a new in-gap state is induced by Zn at low temperatures~\cite{Kimura}. 
This can result in a fast relaxation component in Cu nuclear relaxation. 
Unfortunately, the weight shift between the in-gap state and the original or 
the higher frequency spin susceptibility is not estimated within the existing data. 
In ref.~\cite{Hunt1}, the large distribution function with respect to spin fluctuation
energy is introduced to account consistently 
for a short nuclear spin-lattice relaxation time $T_1$ of $^{139}$La NQR and a moderate $T_1$ of Cu NQR. 
Our model with a short and a long relaxation times is in parallel to such a distribution model.

\section{Conclusion}

We observed Zn-induced wipeout effect 
in addition to already-known strong wipeout effect for underdoped La$_{1.9}$Sr$_{0.1}$CuO$_4$, 
purely Zn-induced wipeout effect for optimally doped La$_{1.85}$Sr$_{0.15}$CuO$_4$, 
but small effect for overdoped La$_{1.8}$Sr$_{0.2}$CuO$_4$. 
The abrupt wipeout effect
without any precursor,  one of the characteristics of the Zn-doping effect, suggests appearance of a spatially
segregated electronic state at low temperatures. 
We associate this strong wipeout effect with the Zn-induced Curie magnetism or its
extended effect of the glassy charge-spin stripe formation. 
Simultaneous magnetic and electric instability plays a key role in
understanding the novel response quite sensitive to nonmagnetic Zn impurity.  

\section{Acknowledgments}
We would like to thanks Y. Koike, C. Panagopoulos, H. Kimura, and K. Hirota for helpful discussions, 
and R. E. Walstedt for critical reading of the manuscript and kind detailed comments.

\begin{figure}
\epsfxsize=2.7in
\epsfbox{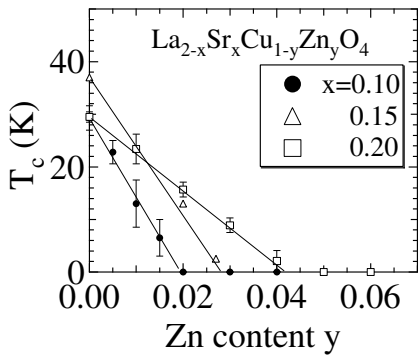}
\vspace{0.0cm}
\caption{
Zn-doping dependence of $T_c$ of La$_{2-x}$Sr$_x$Cu$_{1-y}$Zn$_y$O$_4$ with Sr contents of $x$=0.10, 0.15, and 0.20. 
The amount of the Zn impurity to diminish $T_c$
is estimated to be $y_c$=0.019, 0.028, and 0.042 for $x$=0.10, 0.15, and 0.20, respectively.
}
\label{TcvsZn}
\end{figure}

\vspace{-0.5cm}
\begin{figure}
\epsfxsize=3.2in
\epsfbox{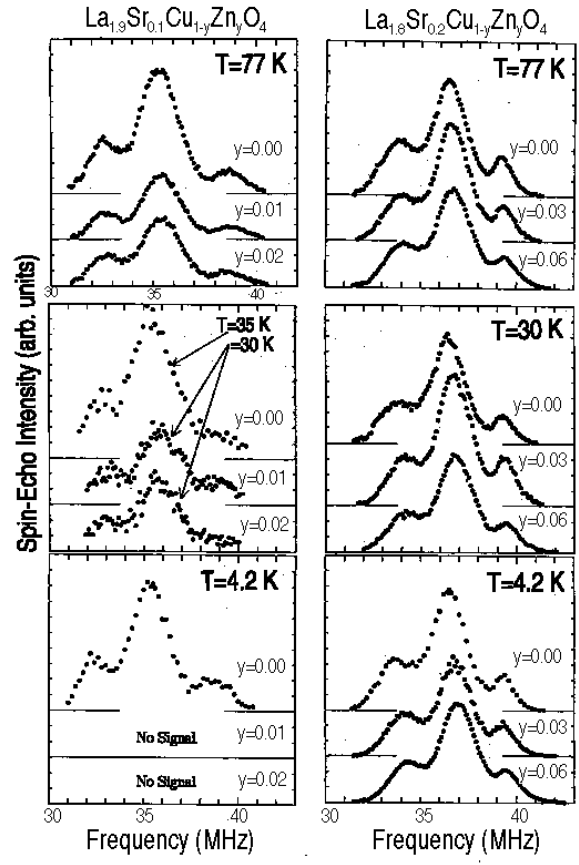}
\vspace{0.3cm}
\caption{
Temperature dependence of Cu NQR frequency spectra of the Zn-doped, Sr-underdoped samples with $x$=0.10 
(Zn contents of $y$=0, 0.01, and
0.02) (left) and of the Zn-doped, Sr-overdoped samples with $x$=0.20 (Zn contents of $y$=0, 0.03, and 0.06) (right), 
at $T$=77 K, 30 or 35 K and 4.2 K (from the top to the bottom). 
The Cu NQR signals for the Zn-doped, Sr-underdoped samples completely diminished at lower temperatures.  
}
\label{NQRspectra}
\end{figure}

\begin{figure}
\epsfxsize=3.3in
\epsfbox{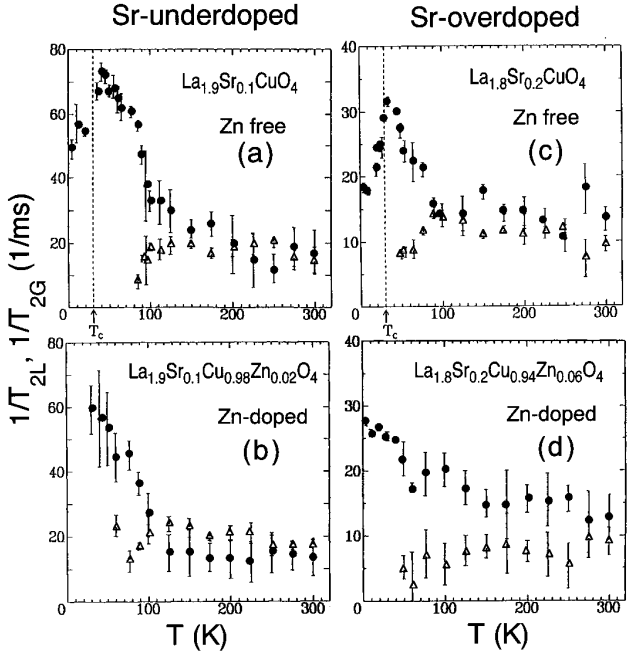}
\vspace{0.1cm}
\caption{
Transverse relaxation rates, Lorentzian decay rates 1/$T_{2L}$ (closed circles) and Gaussian decay rates 1/$T_{2G}$ (open triangles), 
of the Cu nuclear spin-echo decays at each peak frequency for the Sr-underdoped (left) and 
the Sr-overdoped samples (right). 
For the Zn-free superconducting samples in (a) and (c), 
the decrease of the Lorentzian decay rate 1/$T_{2L}$ is observed below $T_c$, irrespective of the wipeout effect. 
For the Zn-doped, Sr-underdoped sample in (b), 
no sizable divergence of 1/$T_{2L}$ nor 1/$T_{2G}$ is observed, when the Cu NQR
signal diminishes at lower temperature. For the Zn-doped, Sr-overdoped sample in (d), 
1/$T_{2L}$ continues to increase below Zn-free $T_c$. 
}
\label{T2}
\end{figure}

\begin{figure}
\epsfxsize=2.2in
\epsfbox{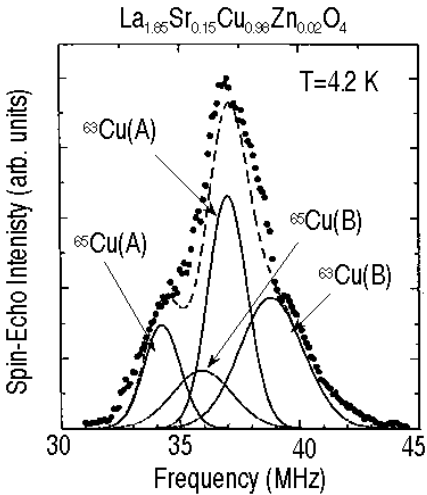}
\vspace{0.2cm}
\caption{
A typical Cu NQR spectrum and the fitted results by the multiple Gaussian functions 
for La$_{1.85}$Sr$_{0.15}$Cu$_{0.98}$Zn$_{0.02}$O$_4$
at 4.2 K. }
\label{Sr15ZnNQR}
\end{figure}

\begin{figure}
\epsfxsize=3.3in
\epsfbox{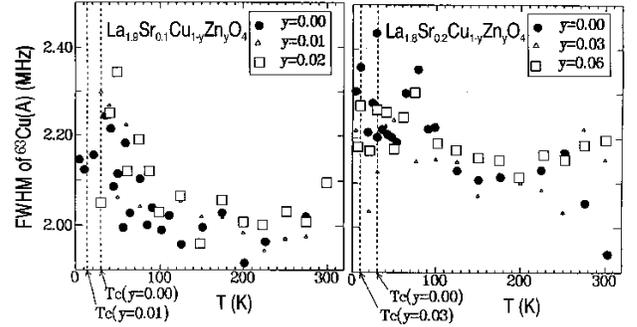}
\vspace{0.5cm}
\caption{
The temperature dependence of the full-width of half maximum of $^{63}$Cu(A). 
Some broadening is seen at low temperatures.  
}
\label{FWHM}
\end{figure}
   
\begin{figure}
\epsfxsize=3.3in
\epsfbox{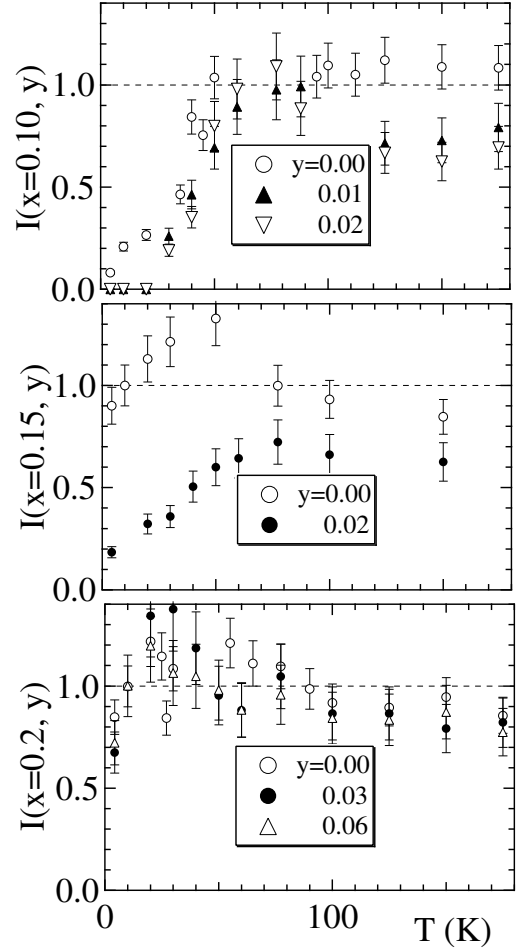}
\vspace{0.0cm}
\caption{
The temperature dependence of the integrated intensity multiplied by temperature 
$I(x$=0.10, $y$), $I(x$=0.15, $y$), and $I(x$=0.2, $y$) for the Zn-free and Zn-doped samples.
}
\label{Intenisty}
\end{figure}

\begin{figure}
\epsfxsize=3.3in
\epsfbox{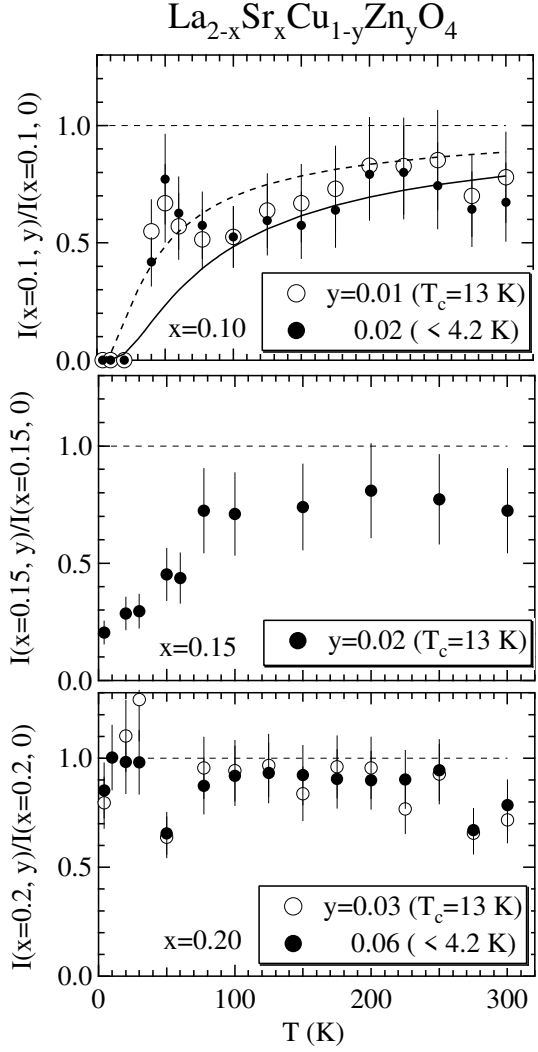}
\vspace{0.0cm}
\caption{
The ratio of the integrated Cu NQR spectra, $I(x$, $y$)/$I(x$, 0) with ($x$=0.10; $y$=0, 0.01, 0.02) (top), 
($$x=0.15; $y$=0, 0.02) (middle) and ($x$=0.20; $y$=0, 0.03, 0.06) (bottom), as functions of temperature. 
The dashed curve is a function of (1-0.01)$^{12\times300/T}$ and the solid
one is (1-0.02)$^{12\times300/T}$. 
The function of (1-$x$)$^{\Sigma n.n.}$ is a probability of finding the nuclei under an assumption that the Cure-type
temperature dependent region around Zn is wiped out 
(the third nearest-neighbor sites to Zn are assumed to be unobservable at $T$=300 K).
}
\label{RelativeIntenisty}
\end{figure}

\begin{figure}
\epsfxsize=3.0in
\epsfbox{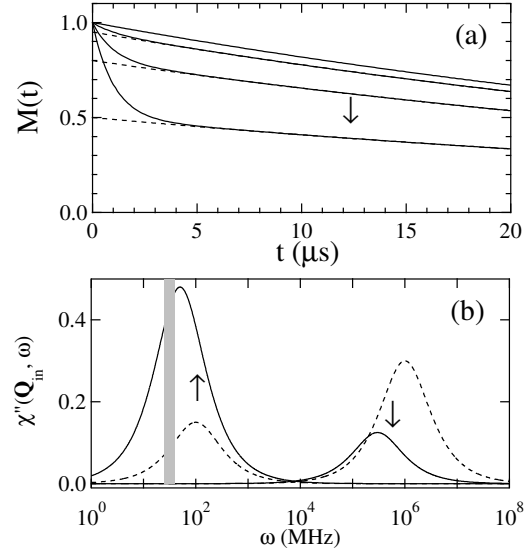}
\vspace{0.0cm}
\caption{
A toy model to describe the full wipeout effect without any precursor below $T_{NQR}$. 
Schematic illustrations of the time spectrum of a muon
spin relaxation $M(t)$ (a) and of the dynamical spin susceptibility 
$\chi"({\bf Q}_{in}, \omega)$ (${\bf Q}_{in}$ is an incommensurate wave vector) (b).  
The arrows indicate cooling down below $T_{NQR}$. 
In Fig. 7(a), the dashed curves are extrapolated from longer relaxation components. 
In Fig. 7(b), the shaded area is an observable NQR frequency region.  
}
\label{ToyModel}
\end{figure}

\end{document}